\documentclass[english,12pt]{article}
\usepackage[T1]{fontenc}
\usepackage[latin1]{inputenc}

\usepackage{geometry}
\usepackage{epsfig}
\usepackage{cite}
\usepackage{color}
\usepackage{comment}
\usepackage{amsmath}
\usepackage{multirow}
\usepackage{bigdelim}

\def\({\left(} 
\def\){\right)}

\begin{document}
\begin{titlepage}
{\begin{flushright}{
 \begin{minipage}{10.5cm}
   KA-TP-28-2012, MAN/HEP/2012/09, MZ-TH/12-27, \\
   CERN-PH-TH/2012-194, IPPP/12/50, DCPT/12/100, \\
   FTUV-12-0709, LPN12-078, TTK-12-32
\end{minipage}}\end{flushright}}
\vspace{0.8cm}
\begin{center} 
  {\Large \bf Release Note -- \textsc{Vbfnlo-2.6.0}} 
\end{center}
\vspace{0.4cm}
\begin{center}
{\renewcommand{\baselinestretch}{4}
K.~Arnold$^{1}$, J.~Bellm$^{1}$, G.~Bozzi$^{2}$, F.~Campanario$^{1}$,
C.~Englert$^{3}$, B.~Feigl$^{1}$, J.~Frank$^{1}$, T.~Figy$^{4}$,
B.~J\"ager$^{5}$,
M.~Kerner$^{1}$, M.~Kubocz$^{6}$, C.~Oleari$^{7}$, S.~Palmer$^{1}$, M.~Rauch$^{1}$, H.~Rzehak$^{8}$, F.~Schissler$^{1}$, O.~Schlimpert$^{1}$,
M.~Spannowsky$^{3}$, D.~Zeppenfeld$^{1}$}
\end{center}
\vspace{0.3cm}
\begin{center}
$^{1}$ Institut f\"{u}r Theoretische Physik, Universit\"at Karlsruhe, Karlsruhe
Institute of Technology, 76128 Karlsruhe,  Germany \\ \noindent
$^{2}$ Dipartimento di Fisica, Universita di Milano and INFN, 20133
Milano, Italy \\ \noindent
$^{3}$ Institute for Particle Physics Phenomenology, University of Durham, Durham, DH1~3LE,
United Kingdom \\ \noindent
$^{4}$ School of Physics and Astronomy, The University of Manchester, Manchester, M13~9PL, United Kingdom\\ \noindent
$^{5}$ Institut f{\"u}r Physik (THEP), Johannes-Gutenberg-Universit{\"a}t, 55099
Mainz, Germany\\ \noindent
$^{6}$ Institut f\"{u}r Theoretische Teilchenphysik und Kosmologie, RWTH Aachen
University, 52056 Aachen, Germany  \\ \noindent
$^{7}$ Dipartimento di Fisica, Universit\`a di Milano-Bicocca and INFN, Sezione di
Milano-Bicocca, 20126 Milano, Italy \\ \noindent
$^{8}$ CERN, CH-1211 Geneva 23, Switzerland (on leave from: Physikalisches Institut Albert-Ludwigs-Universit\"{a}t
Freiburg, Hermann-Herder-Str.\ 3, 79104 Freiburg im Breisgau, Germany) \\
\noindent
\noindent
\end{center}
\vspace{0.4cm}

\begin{abstract}

\textsc{Vbfnlo} is a flexible parton level Monte Carlo program for the
simulation of vector boson fusion (VBF), double and triple vector boson (plus
jet) production in hadronic collisions at next-to-leading order~(NLO) in the
strong coupling constant, as well as Higgs boson plus two jet production via
gluon fusion at the one-loop level.  This note briefly describes the main
additional features and processes that have been added in the new release --
\textsc{Vbfnlo Version 2.6.0}.  At NLO QCD diboson production ($W\gamma$, $WZ$,
$ZZ$, $Z\gamma$ and $\gamma\gamma$), same-sign $W$ pair production via vector
boson fusion and the process $W\gamma\gamma j$ have been implemented (for which
one-loop tensor integrals up to six-point functions are included).  In addition,
gluon induced diboson production can be studied separately at the leading order
(one-loop) level. The diboson processes $WW$, $WZ$ and $W\gamma$ can be run with
anomalous gauge boson couplings, and anomalous couplings between a Higgs and a
pair of gauge bosons is included in $WW$, $ZZ$, $Z\gamma$ and $\gamma\gamma$
diboson production. The code has also been extended to include anomalous gauge
boson couplings for single vector boson production via VBF, and a spin-2 model
has been implemented for diboson pair production via vector boson fusion.
 
\end{abstract}
\vspace{0.3cm}
\today
\end{titlepage}

\section{\textsc{Introduction}}

\textsc{Vbfnlo} \cite{Arnold:2008rz,Arnold:2011wj} is a flexible Monte Carlo
(MC) program for vector boson fusion (VBF), double and triple vector boson (plus
jet) production processes at NLO QCD accuracy. The electroweak corrections to
Higgs boson production via VBF have been included.  In addition, the simulation
of $\cal{CP}$-even and $\cal{CP}$-odd Higgs boson production in gluon fusion,
associated with two additional jets, is implemented at the (one-loop) LO QCD
level.  \textsc{Vbfnlo} can be run in the MSSM, and anomalous couplings of the
Higgs boson and gauge bosons have been implemented for certain processes.
Additionally, two Higgsless extra dimension models are included -- the Warped
Higgsless scenario and a Three-Site Higgsless Model -- for selected processes.

Further information, and the latest version of the code, can be found on the
\textsc{Vbfnlo} webpage
\begin{center}
{\tt http://www-itp.particle.uni-karlsruhe.de/vbfnlo/}
\end{center}
A complete process list is given in Appendix A.

%--------------------------------------------------------------------------------
%================= Section ======================================================
%--------------------------------------------------------------------------------

\section{New processes}

The latest version of \textsc{Vbfnlo} has several new processes implemented at NLO QCD.

\subsection{Same sign $W$ pair production via VBF}

Same sign $W$ pair production with two jets via vector boson fusion
\cite{Jager:2009xx} has been included in \textsc{Vbfnlo 2.6.0}.  This process is
potentially sensitive to new physics signals and, as it gives rise to same sign
dilepton final states, it is also a background to new physics scenarios.  To
distinguish potential signatures of physics beyond the Standard Model from the
effect of higher order corrections, precise theoretical predictions are
essential.  Although the QCD corrections to the integrated cross sections were
found to be relatively small, their effect on several distributions is
appreciable.  When NLO QCD corrections are taken into account, the residual
scale uncertainties are at the 2.5\% level.  The new process IDs are given in
Table~\ref{tab:VVJJ}.

\begin{table}[htb!]
\newcommand{\lstrut}{{$\strut\atop\strut$}}
\begin{center}
\small
\begin{tabular}{c|l}
\hline
&\\
\textsc{ProcId} & \textsc{Process}\\
\hline
&\\
\bf 250 & $p \overset{\mbox{\tiny{(--)}}}{p} \to W^{+}W^{+} \,  jj\to \ell_{1}^{+} \nu_{\ell_{1}} \ell_{2}^{+} \nu_{\ell_{2}} \, jj$  \\
\bf 260 & $p \overset{\mbox{\tiny{(--)}}}{p} \to W^{-}W^{-} \,  jj\to \ell_{1}^{-} \bar{\nu}_{\ell_{1}} \ell_{2}^{-} \bar{\nu}_{\ell_{2}} \, jj$  \\
&\\
\hline
\end{tabular}
\caption {\em  New process IDs for diboson + 2 jet
production via vector boson fusion at NLO QCD accuracy.}
\vspace{0.2cm}
\label{tab:VVJJ}
\end{center}
\end{table}

\subsection{Diboson production and gluon-induced contributions}

A good understanding of diboson processes at the LHC is essential, as not only
do they allow the study of the Standard Model's gauge structure, but they also
provide a background to Higgs boson and new physics searches.  Anomalous triple
gauge boson couplings are included in the $WZ$ and $W\gamma$
processes\footnote{As in the rest of \textsc{Vbfnlo}, no neutral triple  gauge
couplings are included.}.  (For contributions from anomalous $HVV$ couplings in
gluon induced processes -- Fig.~\ref{fig:ggVV} -- see below.) The new processes
are included at NLO QCD accuracy under the process IDs shown in
Table~\ref{tab:diboson}.
\begin{table}[htb!]
\newcommand{\lstrut}{{$\strut\atop\strut$}}
\begin{center}
\small
\begin{tabular}{c|l|l}
\hline
&\\
\textsc{ProcId} & \textsc{Process} & \textsc{BSM}  \\
&\\
\hline
&\\
\bf 310 & $p \overset{\mbox{\tiny{(--)}}}{p} \to W^{+}Z \to  \ell_{1}^{+} \nu_{\ell_1}  \ell_{2}^{+} \ell_{2}^{-} $ & \ldelim \} {2}{0.8cm} \multirow{2}{*}{anomalous $VVV$ couplings}\\
\bf 320 & $p \overset{\mbox{\tiny{(--)}}}{p} \to W^{-}Z \to \ell_{1}^{-} \bar{\nu}_{\ell_{1}}  \ell_{2}^{+} \ell_{2}^{-} $ & \\
\bf 330 & $p \overset{\mbox{\tiny{(--)}}}{p} \to ZZ \to \ell_{1}^{-} \ell_{1}^{+}  \ell_{2}^{-} \ell_{2}^{+} $ & anomalous $HVV$ couplings\\
\bf 340 & $p \overset{\mbox{\tiny{(--)}}}{p} \to W^{+}\gamma \to \ell_{1}^{+} \nu_{\ell_1} \gamma $ & \ldelim \} {2}{0.8cm} \multirow{2}{*}{anomalous $VVV$ couplings}\\
\bf 350 & $p \overset{\mbox{\tiny{(--)}}}{p} \to W^{-}\gamma \to \ell_{1}^{-} \bar{\nu}_{\ell_1} \gamma $ & \\
\bf 360 & $p \overset{\mbox{\tiny{(--)}}}{p} \to Z\gamma \to \ell_{1}^{-} \ell_{1}^{+}  \gamma $ & \ldelim \} {2}{0.8cm} \multirow{2}{*}{anomalous $HVV$ couplings}\\
\bf 370 & $p \overset{\mbox{\tiny{(--)}}}{p} \to \gamma\gamma $ & \\
&\\
\hline
\end{tabular}
\caption {\em  Process IDs for the new diboson production processes at NLO
  QCD accuracy.}
\vspace{0.2cm}
\label{tab:diboson}
\end{center}
\end{table}

In addition to the NLO QCD corrections, the gluon-induced fermionic loop
processes can be included in those processes with neutral final states (including
$WW$ production, which was already implemented in \textsc{Vbfnlo}, with process
ID~300).  Both continuum production via box diagrams as well as production via an s-channel
Higgs boson resonance (shown in Fig.~\ref{fig:ggVV}) are available, with full interference effects.  
\begin{figure}[!htb]
\begin{center}
         \resizebox{0.75\hsize}{!}{\includegraphics*{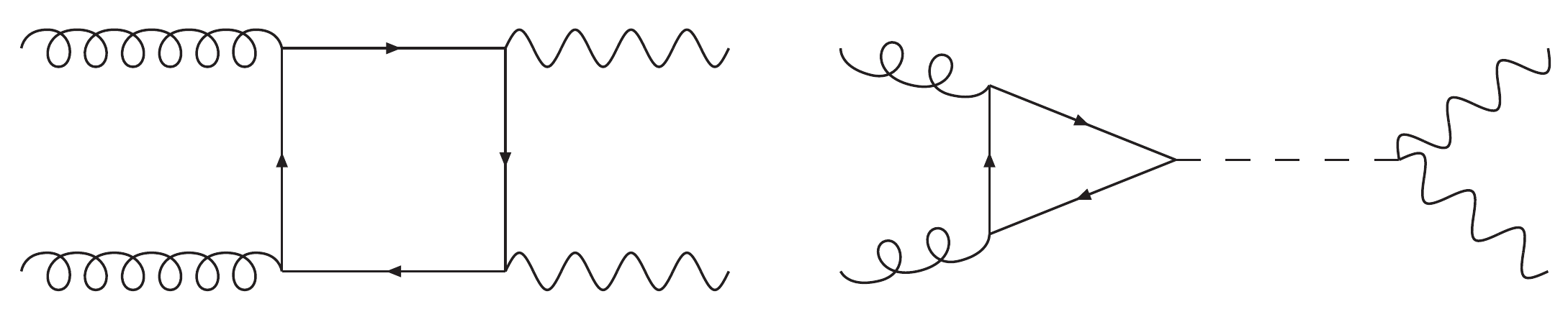}} 
\end{center}
     \caption{Gluon-induced contributions to diboson production.}
     \label{fig:ggVV}
\end{figure} 

The {\tt vbfnlo.dat} flag {\tt FERMIONLOOP} controls these contributions:
  \begin{itemize}
  \item {\tt 0} switches off these processes
  \item {\tt 1} includes only the box contribution 
  \item {\tt 2} includes only diagrams via an s-channel Higgs resonance
  \item {\tt 3} includes both contributions including interference effects. 
  \end{itemize}
Although these contributions are formally of higher order, their effect can
still be significant.  Anomalous couplings between a Higgs boson and a pair of
gauge bosons can be included in those processes with neutral final states.  The
gluon-induced loop diagrams can be studied separately in the new release of
\textsc{Vbfnlo} using the process IDs of Table~\ref{tab:gluon_diboson} and the
executable {\tt ggflo}.
\begin{table}[htb!]
\newcommand{\lstrut}{{$\strut\atop\strut$}}
\begin{center}
\small
\begin{tabular}{c|l|l}
\hline
&\\
\textsc{ProcId} & \textsc{Process} & \textsc{BSM}  \\
&\\
\hline
&\\
\bf 4300 & $gg \to W^{+}W^{-} \to \ell_{1}^{+} \nu_{\ell_{1}} \ell_{2}^{-}\bar{\nu}_{\ell_{2}} $ & \ldelim \} {4}{0.8cm} \multirow{4}{*}{anomalous $HVV$ couplings}\\
\bf 4330 & $gg \to ZZ \to \ell_{1}^{-} \ell_{1}^{+}  \ell_{2}^{-} \ell_{2}^{+} $ & \\
\bf 4360 & $gg \to Z\gamma \to \ell_{1}^{-} \ell_{1}^{+}  \gamma $ & \\
\bf 4370 & $gg \to \gamma\gamma $ & \\
&\\
\hline
\end{tabular}
\caption {\em  Process IDs for gluon induced diboson production at LO (one-loop) QCD.}
\vspace{0.2cm}
\label{tab:gluon_diboson}
\end{center}
\end{table}

\subsection{Triple vector boson production in association with a hadronic jet}

Finally, the triboson plus jet processes given in Table~\ref{tab:tribosonjet}
have also been included at NLO QCD level\footnote{These
processes are disabled by default and must be enabled at compilation using the
{\tt configure} option {\tt ----enable-processes=all} or {\tt
----enable-processes=tribosonjet}.} \cite{Campanario:2011ud}.  As a process involving multiple
electroweak bosons and jets, this is an important channel in which to compare
experimental data with the predictions of the Standard Model. The NLO QCD
corrections to the total cross section are sizeable, and have a non-trivial
phasespace dependence.

\begin{table}[htb!]
\begin{center}
\small
\begin{tabular}{c|l}
\hline
&\\
\textsc{ProcId} & \textsc{Process} \\
&\\
\hline
&\\
\bf 800 & $p \overset{\mbox{\tiny{(--)}}}{p}  \to W^{+} \gamma \gamma j  \to \ell^{+} \nu_{\ell} \gamma \gamma j $ \\
\bf 810 & $p \overset{\mbox{\tiny{(--)}}}{p}  \to W^{-} \gamma \gamma j \to \ell^{-} \bar \nu_{\ell} \gamma \gamma j $  \\
& \\
\hline
\end{tabular}
\caption{ \em  Process IDs for triboson production in association with a hadronic jet at
NLO QCD.}
\vspace{0.2cm}
\label{tab:tribosonjet}
\end{center}
\end{table}

%--------------------------------------------------------------------------------
%================= Section ======================================================
%--------------------------------------------------------------------------------

\section{New features}

In addition to the new processes described above, several existing procedures
have been extended and new features added.

\subsection{Tensor reduction routines}
The tensor reduction routines for up to 4 external legs have been
extended for general kinematics. Three and four point tensor integrals are
extended to deal with Rank 3 and Rank 4 integrals, respectively. 
Furthermore, one-loop tensor reduction routines for up to 6 external legs have been
included for the massless case.  The  tensor integrals are
implemented following Ref.~\cite{Campanario:2011cs} and can be found in the directory {\tt loops/TenRed}.

\subsection{Anomalous couplings}
Anomalous gauge boson couplings have been implemented for single vector boson
production via VBF, which have been seen to have an effect on some distributions
\cite{Jager:2010aj}, such as the azimuthal angle separation of the tagging jets.
The existing diboson process $pp \rightarrow WW$ has been modified to include
anomalous $VVV$ couplings, as well as anomalous $HVV$ couplings in the
gluon-induced contributions.  The relevant process IDs are given in
Table~\ref{tab:anom}. To run \textsc{Vbfnlo} with anomalous couplings, the
switch {\tt ANOM\_CPL} in the input file {\tt vbfnlo.dat} must be switched to
{\tt true}.  The anomalous coupling parameters are then input via {\tt
anomV.dat} or (for $HVV$ couplings)
{\tt anom\_HVV.dat}.
\begin{table}[htb!]
\newcommand{\lstrut}{{$\strut\atop\strut$}}
\begin{center}
\small
\begin{tabular}{c|l|l}
\hline
&\\
\textsc{ProcId} & \textsc{Process} & \textsc{Bsm}\\
\hline
&\\
\bf 120 & $p \overset{\mbox{\tiny{(--)}}}{p} \to Z \, jj \to \ell^{+} \ell^{-} \, jj$ & \ldelim \} {6}{0.8cm} \multirow{5}{*}{anomalous  couplings} \\
\bf 121 & $p \overset{\mbox{\tiny{(--)}}}{p} \to Z  \, jj\to \nu_\ell \bar{\nu}_\ell \, jj$ & \\
\bf 130 & $p \overset{\mbox{\tiny{(--)}}}{p} \to W^{+} \,  jj\to \ell^{+} \nu_\ell \, jj$ & \\
\bf 140 & $p \overset{\mbox{\tiny{(--)}}}{p} \to W^{-} \, jj\to \ell^{-} \bar{\nu}_\ell  \, jj$ &\\
\bf 150 & $p \overset{\mbox{\tiny{(--)}}}{p} \to \gamma \, jj$ &\\
&\\
\hline
&\\
\bf 300 & $p \overset{\mbox{\tiny{(--)}}}{p} \to W^{+}W^{-} \to \ell_{1}^{+} \nu_{\ell_{1}} \ell_{2}^{-}\bar{\nu}_{\ell_{2}} $ & anomalous $VVV$ and $HVV$ couplings\\
&\\
\hline
\end{tabular}
\caption {\em  Process IDs for existing processes which have been extended to include anomalous couplings.}
\vspace{0.2cm}
\label{tab:anom}
\end{center}
\end{table}

\subsection{Spin-2 model}
A spin-2 model has been implemented, using an effective Lagrangian to describe
the interactions of spin-2 particles with electroweak gauge bosons for two
cases: an isospin singlet spin-2 state and a spin-2 triplet, as described in
Ref.~\cite{frank}.  This spin-2 model is implemented for diboson plus two jets
production via vector boson fusion ($pp \rightarrow W^{+}W^{-}jj$, $pp \rightarrow ZZjj$
and $pp \rightarrow WZjj$, process IDs 200 - 230).  For these processes a spin-2
resonance is included in addition to the Standard Model diagrams (i.e.\ both
Higgs boson and spin-2 diagrams are calculated).  

The file {\tt spin2coupl.dat} is used to set the parameters for the spin-2
models.  It is read if the switch {\tt SPIN2} in {\tt vbfnlo.dat} is set to {\tt
true}, and will only run if the spin-2 models were enabled at compilation
using the {\tt configure} option {\tt ----enable-spin2}.

For the singlet spin-2 field, $T^{\mu\nu}$, the effective Lagrangian is
\begin{equation}
 \mathcal{L}_{\text{singlet}} = \frac{1}{\Lambda} T_{\mu \nu} \left( f_{1} B^{\alpha \nu} B^{\mu}_{\alpha} + f_{2} W_{i}^{\alpha \nu} W^{i,\mu}_{\alpha} + f_3 \widetilde{B}^{\alpha \nu} B^\mu_{\alpha} 
+f_4\widetilde{W}_i^{\alpha \nu} W^{i, \mu}_{\alpha} +  2 f_{5} (D^{\mu} \Phi)^{\dagger} (D^{\nu}\Phi) \right),
\end{equation}

and for the spin-2 triplet field, $T_{j}^{\mu\nu}$, the effective Lagrangian is given by
\begin{equation}
 \mathcal{L}_{\text{triplet}} = \frac{1}{\Lambda} T_{\mu\nu j} \left( f_{6} (D^{\mu}\Phi)^{\dagger} \sigma^{j} (D^{\nu} \Phi) + f_{7} W^{j,\mu}_{\alpha} B^{\alpha\nu} \right),
\end{equation}
where $W$ and $B$ are the usual electroweak field strength tensors,
$\widetilde{W}$ and $\widetilde{B}$ the dual field strength tensors, $\Phi$ is
the Higgs field and $D^{\mu}$ is the covariant derivative.  $f_{i}$ are variable
coupling parameters and $\Lambda$ is the characteristic energy scale of the new
physics.

In order to preserve unitarity, a formfactor is introduced to multiply the
amplitudes.  The formfactor has the form:
\begin{equation}
 f(q_{1}^{2}, q_{2}^{2}, p^{2}_{\text{sp2}}) = \left( \frac{\Lambda_{ff}^{2}} {\left|q_{1}^{2}\right| + \Lambda_{ff}^{2}} \cdot \frac{\Lambda_{ff}^{2}} {\left|q_{2}^{2}\right| + \Lambda_{ff}^{2}} \cdot \frac{\Lambda_{ff}^{2}} {\left|p_{\text{sp2}}^{2}\right| + \Lambda_{ff}^{2}} \right) ^{n_{ff}}.
\end{equation}
Here $p^{2}_{\text{sp2}}$ is the invariant mass of a virtual s-channel spin-2 particle
and $q_{1,2}^{2}$ are the invariant masses of the electroweak bosons.  The
energy scale $\Lambda_{ff}$ and the exponent $n_{ff}$ describe the scale of the
cutoff and the suppression power.

The input parameters used by \textsc{Vbfnlo} are
\begin{itemize}
 \item {\tt F1,F2,F3,F4,F5:} Coupling parameters for the spin-2 singlet field. 
Default values are  {\tt F1=F2=F5=1, F3=F4=0}.
 \item {\tt F6,F7:} Coupling parameters for the spin-2 triplet field.  Default
values are set to {\tt 1}.
 \item {\tt LAMBDA:} Energy scale of the couplings in GeV.  Default value is 
{\tt 1500}~GeV.
 \item {\tt LAMBDAFF:} Energy scale of the formfactor in GeV.  Default value is 
{\tt 3000}~GeV.
 \item {\tt NFF:} Exponent of the formfactor.  Default value is {\tt 4}.
\end{itemize}
Note that a graviton corresponds to {\tt F1=F2=F5=1} and {\tt F3=F4=F6=F7=0}.

\textsc{Vbfnlo} also needs the masses and branching ratios (into SM gauge bosons) of the spin-2
particles.  
\begin{itemize}
 \item {\tt SP2MASS:} Mass of the spin-2 singlet particle in GeV.  Default value is 
{\tt 1000}~GeV.
 \item {\tt MSP2TRIPPM:} Mass of charged spin-2 triplet particles in GeV.  Default 
value is {\tt 1000}~GeV.
 \item {\tt MSP2TRIPN:} Mass of neutral spin-2 triplet particle in GeV.  Default 
value is {\tt 1000}~GeV.
 \item {\tt BRRAT:} Branching ratio into SM gauge bosons for spin-2 singlet particle.  Default value is 
{\tt 1}.
 \item {\tt BRRATTRIPPM:} Branching ratio into SM gauge bosons for charged spin-2 triplet particles.  
Default value is~{\tt 1}.
 \item {\tt BRRATTRIPN:} Branching ratio into SM gauge bosons for neutral spin-2 triplet particle.  Default 
value is~{\tt 1}.
\end{itemize}

A new process $pp \rightarrow \gamma \gamma jj$ (Table \ref{tab:aajj}) has been
added which includes only the resonant spin-2 diagrams -- this can be compared
to the existing process of Higgs boson production via VBF, where the Higgs
decays into photons (process ID~101).  The default values given above are
intended for processes 200-230. Corresponding values for light spin-2 resonances
in process 191 can be found in Ref.~\cite{frank}\footnote{Note that in
Ref.~\cite{frank} this process is referred to as process 240.}. 

\begin{table}[htb!]
\newcommand{\lstrut}{{$\strut\atop\strut$}}
\begin{center}
\small
\begin{tabular}{c|l|l}
\hline
&\\
\textsc{ProcId} & \textsc{Process} & \textsc{BSM}\\
\hline
&\\
\bf 191 & $p \overset{\mbox{\tiny{(--)}}}{p} \to S_{2} \,jj \to \gamma\gamma \, jj$ & spin-2 resonant production only \\
&\\
\hline
\end{tabular}
\caption {\em  Process ID~for production of a spin-2 particle $S_{2}$ with 2 jets
 via vector boson fusion at NLO QCD accuracy.}
\vspace{0.2cm}
\label{tab:aajj}
\end{center}
\end{table}

\subsection{Histograms}

The error calculation for the real emission output can now be controlled via
the input file {\tt histograms.dat}. \textsc{Vbfnlo} can calculate the Monte
Carlo error for each bin and output this to the raw histogram data output for 1D
and 2D histograms. For the gnuplot histogram output only the 1D histograms can
display the error bars\footnote{Error calculation is not implemented for the
other (\textsc{Root} or \textsc{TopDrawer}) histogram formats.}.

\begin{itemize}
 \item {\tt CALC\_ERROR\_GNUPLOT:} Enable or disable y-error bars in 1D gnuplot histograms.
                                 Default is {\tt false}.
 \item {\tt CALC\_ERROR\_1D:} Enable or disable y-error bars in raw 1D histogram output.
                            Default is {\tt true}.
 \item {\tt CALC\_ERROR\_2D:} Enable or disable z-error bars in raw 2D histogram output.
                            Default is {\tt false}.
\end{itemize}

Furthermore, \textsc{Vbfnlo} uses a smearing between adjacent bins to avoid
artefacts at NLO when the real emission kinematics and the corresponding
subtraction term fall into different bins. As this can lead to remnants at the sharp
edges caused by cuts, the smearing can be switched off.
\begin{itemize}
 \item {\tt SMEARING:} Enable or disable smearing.  Default is {\tt true}.
 \item {\tt SMEAR\_VALUE:} Set the bin fraction where the bin smearing is active.
                          The part that is put to the next bin becomes larger when
                          the x-value is closer to a bin border. Default is {\tt 0.2}.
\end{itemize}

\subsection{SUSY options for electroweak corrections}

When running \textsc{Vbfnlo} in the MSSM, it is now possible to set the mass of
the Higgs bosons in the electroweak loops to either their tree-level value or their
corrected value, using the flag {\tt MH\_LOOPS} in {\tt susy.dat}.  

In some areas of the MSSM parameter space, the electroweak loop corrections can
be the dominant contribution to the cross section.  In this case, the squared
electroweak corrections from the (s)fermion corrections are important and can be
included in \textsc{Vbfnlo} using the {\tt susy.dat} flag {\tt LOOPSQR\_SWITCH}.
If set to {\tt true} the amplitude is given by
\begin{equation}
  |\mathcal{M}_{\text{Born}}|^{2} + 2 \Re\left[\mathcal{M}^{*}_{\text{Born}} \mathcal{M}_{loop} \right] + |\mathcal{M}_{\text{(s)fermion loop}}|^{2} .
\end{equation}
Note that the loop squared component is only added if
$|\mathcal{M}_{\text{(s)fermion loop}}|$ is greater than 10\% of
$|\mathcal{M}_{\text{Born}}|$.

\section{Other changes}

Since the previous release, \textsc{Version 2.5.0}, some changes have been made
that alter previous results (events, cross sections and distributions).

\subsection{Allowed virtuality of resonance}
In the phasespace generators, the allowed range of the virtuality of a resonance
of intermediate vector bosons has been increased.  This mainly affects processes
where an intermediate $Z$ boson decays into a pair of neutrinos -- i.e.
\begin{itemize}
 \item \ $pp \rightarrow Hjj \rightarrow ZZjj \rightarrow \ell^{+} \ell^{-} \nu \overline{\nu} jj$ via vector boson fusion (process ID 107) and gluon fusion (process ID 4107)
 \item $pp \rightarrow Hjjj \rightarrow ZZjjj \rightarrow \ell^{+} \ell^{-} \nu \overline{\nu} jjj$ (process ID 117)
 \item $pp \rightarrow H\gamma jj \rightarrow ZZ\gamma jj \rightarrow \ell^{+} \ell^{-} \nu \overline{\nu} \gamma jj$ (process ID 2107)
 \item $pp \rightarrow ZZjj \rightarrow \ell^{+} \ell^{-} \nu \overline{\nu} jj$ (process ID 211)
\end{itemize}
This not only affects the cross sections
for these processes, but also means that the events produced by
\textsc{Vbfnlo-2.6.0} will differ from those produced by \textsc{Vbfnlo-2.5},
even if the same random numbers are used.

\subsection{Matrix element $H \rightarrow ZZ \rightarrow 4\ell$}
A bug was found and fixed in the implementation of the matrix element
calculating the decay $H \rightarrow ZZ \rightarrow 4\ell$.  

\subsection{Anomalous couplings}
Several changes have been made to the implementation of the anomalous couplings.
 For Higgs production via vector boson fusion (process IDs 100-107) the variable
{\tt TREEFAC}, which multiplies the Standard Model contribution to the
tree-level $HVV$ couplings, has been corrected and altered -- now, separate
factors for $HZZ$ and $HWW$ are input ({\tt TREEFACZ} and {\tt TREEFACW}
respectively). 

When working with anomalous $HVV$ couplings two types of formfactors can be
applied which model effective, momentum dependent $HVV$ vertices, motivated by
new physics entering with a large scale $\Lambda$ at loop level.  Corrections to
the $HVV$ formfactor $F_{2}$, where
\begin{eqnarray}
\label{eq:ff2}
F_2 &=& -2 \,\Lambda^2 \, C_0\!\left(q_1^2, q_2^2, (q_1+q_2)^2,
\Lambda^2\right). 
\end{eqnarray}
(where the $q_i$ are the momenta of the vector bosons and $C_{0}$ is the scalar
one-loop three point function in the notation of Ref.\cite{Passarino:1978jh})
have been made.   The implementation of the parametrization described by {\tt
PARAMETR3} -- where the input determining the anomalous couplings is in terms of
the dimension-6 operators ($\mathcal{O}_{W}$, $\mathcal{O}_{B}$,
$\mathcal{O}_{WW}$ and $\mathcal{O}_{BB}$, see
\cite{Hagiwara:1993qt,Hagiwara:1993ck}) -- has also been corrected. 

If anomalous triple (and quartic) gauge boson couplings are being studied, a formfactor, given by
\begin{equation}
 F = \left(1 + \frac{s}{\Lambda^{2}} \right)^{-p},
\end{equation}
can be applied in order to preserve unitarity, where $\Lambda$ is again the
scale of new physics.  The momentum dependence of an applied formfactor (i.e.\
$s$) is now universal for each phasespace point, with the invariant mass of the
bosons as the scale.  This ensures the proper cancellations for anomalous
contributions and affects both the cross sections and the
distributions significantly.

The values of the formfactor scale $\Lambda$ and suppression $p$
can be set to different values for each input describing the triboson couplings.
 In the parameterization {\tt TRIANOM = 2} of the L3 Collaboration
\cite{Achard:2004kn}, the formfactor scales for $\Delta \kappa_{\gamma}$ and
$\Delta \kappa_{Z}$ are now separately set, and the consistency of related
parameters (i.e.\ $\Delta g_{1}^{Z}$, $\Delta \kappa_{\gamma}$ and $\Delta
\kappa_{Z}$) is enforced when formfactors are applied.

When processes involving resonant Higgs diagrams (e.g.\ $WWW$
production) are studied with anomalous couplings, the Higgs width is now
calculated with the appropriate anomalous $HVV$ couplings (the anomalous $HVV$
couplings in the production amplitudes were taken into account in previous
versions of \textsc{Vbfnlo}).  Various corrections have also been made to the
anomalous triboson couplings in diboson plus jet processes (these were
incorporated into the intermediate release \textsc{Vbfnlo} 2.5.3).

\subsection{VBF Higgs boson production in association with three jets}
A small bug was found and fixed in the calculation of the processes $pp \rightarrow H jjj$, with process IDs 110-117.

%--------------------------------------------------------------------------------
%================= Section ======================================================
%--------------------------------------------------------------------------------

\section*{Acknowledgments}

We are grateful to Simon Pl\"atzer, Manuel B\"ahr, Martin Brieg, Florian Geyer, Nicolas Greiner,
Christoph Hackstein, Vera Hankele, Gunnar Kl\"amke, Stefan Prestel and
Malgorzata Worek for their past contributions to the \textsc{Vbfnlo} code.  We
would also like to thank Julien Baglio for help in testing the code.  TF would
like to thank the North American Foundation for The University of Manchester and
George Rigg for their financial support.

\providecommand{\href}[2]{#2}

\newpage
\section*{Appendix A: Process List}
\label{app:proc_list}
The following is a complete list of all processes available in \textsc{Vbfnlo},
including any Beyond the Standard Model (BSM) effects that are implemented. 
Firstly, the processes that are accessed via the {\tt vbfnlo} executable are given.

\begin{table}[htb!]
\newcommand{\lstrut}{{$\strut\atop\strut$}}
\begin{center}
\small
\begin{tabular}{c|l|l}
\hline
&\\
\textsc{ProcId} & \textsc{Process} & \textsc{Bsm} \\
&\\
\hline
&\\
\bf 100 & $p \overset{\mbox{\tiny{(--)}}}{p} \to H \, jj$ & \ldelim \} {10}{0.8cm} \multirow{10}{*}{anomalous HVV couplings, MSSM} \\
\bf 101 & $p \overset{\mbox{\tiny{(--)}}}{p} \to H \, jj\to \gamma\gamma \, jj$ & \\
\bf 102 & $p \overset{\mbox{\tiny{(--)}}}{p} \to H \, jj\to \mu^+\mu^- \, jj$ & \\
\bf 103 & $p \overset{\mbox{\tiny{(--)}}}{p} \to H \, jj\to \tau^+\tau^- \, jj$ & \\
\bf 104 & $p \overset{\mbox{\tiny{(--)}}}{p} \to H \, jj\to b\bar{b} \, jj$ & \\
\bf 105 & $p \overset{\mbox{\tiny{(--)}}}{p} \to H \, jj\to W^{+}W^{-} \, jj\to \ell_{1}^+\nu_{\ell_{1}} \ell_{2}^- 
\bar{\nu}_{\ell_{2}} \,jj$ &  \\
\bf 106 & $p \overset{\mbox{\tiny{(--)}}}{p} \to H \, jj\to ZZ \, jj\to \ell_{1}^+ \ell_{1}^- \ell_{2}^+ 
\ell_{2}^- \,jj$ & \\
\bf 107 & $p \overset{\mbox{\tiny{(--)}}}{p} \to H \, jj\to ZZ \, jj\to \ell_{1}^+ \ell_{1}^- \nu_{\ell_{2}}  
\bar{\nu}_{\ell_{2}} \,jj$ & \\
&\\
\hline
&\\
\bf 110 & $p \overset{\mbox{\tiny{(--)}}}{p} \to H \, jjj$ \\
\bf 111 & $p \overset{\mbox{\tiny{(--)}}}{p} \to H \, jjj\to \gamma\gamma \, jjj$ \\
\bf 112 & $p \overset{\mbox{\tiny{(--)}}}{p} \to H \, jjj\to \mu^+\mu^- \, jjj$ \\
\bf 113 & $p \overset{\mbox{\tiny{(--)}}}{p} \to H \, jjj\to \tau^+\tau^- \, jjj$ \\
\bf 114 & $p \overset{\mbox{\tiny{(--)}}}{p} \to H \, jjj\to b\bar{b} \, jjj$ \\
\bf 115 & $p \overset{\mbox{\tiny{(--)}}}{p} \to H \, jjj\to W^+W^- \, jjj\to \ell_{1}^{+}\nu_{\ell_{1}} \ell_{2}^- 
\bar{\nu}_{\ell_{2}} \,jjj$ \\
\bf 116 & $p \overset{\mbox{\tiny{(--)}}}{p} \to H \, jjj\to ZZ \, jjj\to \ell_{1}^+ \ell_{1}^- \ell_{2}^+ \ell_{2}^- \,jjj$ \\
\bf 117 & $p \overset{\mbox{\tiny{(--)}}}{p} \to H \, jjj\to ZZ \, jjj\to \ell_{1}^+ \ell_{1}^- \nu_{\ell_{2}}  
\bar{\nu}_{\ell_{2}} \,jjj$ \\
&\\
\hline
&\\
\bf 2100 & $p \overset{\mbox{\tiny{(--)}}}{p} \to H \gamma \, jj$  \\
\bf 2101 & $p \overset{\mbox{\tiny{(--)}}}{p} \to H \gamma \, jj\to \gamma\gamma \gamma \, jj$ \\
\bf 2102 & $p \overset{\mbox{\tiny{(--)}}}{p} \to H \gamma \, jj\to \mu^+\mu^- \gamma \, jj$ \\
\bf 2103 & $p \overset{\mbox{\tiny{(--)}}}{p} \to H \gamma \, jj\to \tau^+\tau^- \gamma \, jj$ \\
\bf 2104 & $p \overset{\mbox{\tiny{(--)}}}{p} \to H \gamma \, jj\to b\bar{b} \gamma \, jj$ \\
\bf 2105 & $p \overset{\mbox{\tiny{(--)}}}{p} \to H \gamma \, jj\to W^+W^- \gamma \, jj\to \ell_{1}^+\nu_{\ell_{1}} \ell_{2}^- 
\bar{\nu}_{\ell_{2}} \gamma \,jj$ \\
\bf 2106 & $p \overset{\mbox{\tiny{(--)}}}{p} \to H \gamma \, jj\to ZZ \gamma \, jj\to \ell_{1}^+ \ell_{1}^- \ell_{2}^+ 
\ell_{2}^- \gamma \,jj$ \\
\bf 2107 & $p \overset{\mbox{\tiny{(--)}}}{p} \to H \gamma \, jj\to ZZ \gamma \, jj\to \ell_{1}^+ \ell_{1}^- \nu_{\ell_{2}}  
\bar{\nu}_{\ell_{2}} \gamma \,jj$ \\
&\\
\hline
\end{tabular}
\end{center}
\end{table}

\newpage

\begin{table}[htb!]
\newcommand{\lstrut}{{$\strut\atop\strut$}}
\begin{center}
\small
\begin{tabular}{c|l|l}
\hline
&\\
\textsc{ProcId} & \textsc{Process} & \textsc{Bsm}\\
\hline
&\\
\bf 120 & $p \overset{\mbox{\tiny{(--)}}}{p} \to Z \, jj \to \ell^{+} \ell^{-} \, jj$ & \ldelim \} {6}{0.8cm} \multirow{5}{*}{anomalous  couplings} \\
\bf 121 & $p \overset{\mbox{\tiny{(--)}}}{p} \to Z  \, jj\to \nu_\ell \bar{\nu}_\ell \, jj$ & \\
\bf 130 & $p \overset{\mbox{\tiny{(--)}}}{p} \to W^{+} \,  jj\to \ell^{+} \nu_\ell \, jj$ & \\
\bf 140 & $p \overset{\mbox{\tiny{(--)}}}{p} \to W^{-} \, jj\to \ell^{-} \bar{\nu}_\ell  \, jj$ &\\
\bf 150 & $p \overset{\mbox{\tiny{(--)}}}{p} \to \gamma \, jj$ &\\
&\\
\hline
&\\
\bf 191 & $p \overset{\mbox{\tiny{(--)}}}{p} \to S_{2} \, jj\to \gamma\gamma \, jj$ & only spin-2 resonant production\\
&\\
\hline
&\\
\bf 200 & $p \overset{\mbox{\tiny{(--)}}}{p} \to W^{+}W^{-} \, jj \to \ell_{1}^{+} \nu_{\ell_{1}} \ell_{2}^{-}
\bar{\nu}_{\ell_{2}} \, jj$ & anomalous couplings, Kaluza-Klein \& spin-2 models\\
\bf 210 & $p \overset{\mbox{\tiny{(--)}}}{p} \to ZZ  \, jj\to \ell_{1}^{+} \ell_{1}^{-} \ell_{2}^{+} \ell_{2}^{-} \, jj$ & \ldelim \} {5}{0.8cm} \multirow{5}{*}{Kaluza-Klein models, spin-2 models}\\
\bf 211 & $p \overset{\mbox{\tiny{(--)}}}{p} \to ZZ  \, jj\to \ell_{1}^{+} \ell_{1}^{-} \nu_{\ell_{2}} \bar{\nu}_{\ell_{2}} \, jj$ & \\
\bf 220 & $p \overset{\mbox{\tiny{(--)}}}{p} \to W^{+}Z \,  jj\to \ell_{1}^{+} \nu_{\ell_{1}} \ell_{2}^{+} \ell_{2}^{-} \, jj$ & \\
\bf 230 & $p \overset{\mbox{\tiny{(--)}}}{p} \to W^{-}Z \, jj\to \ell_{1}^{-} \bar{\nu}_{\ell _{1}} \ell_{2}^{+} \ell_{2}^{-} \, jj$ & \\
\bf 250 & $p \overset{\mbox{\tiny{(--)}}}{p} \to W^{+}W^{+} \,  jj\to \ell_{1}^{+} \nu_{\ell_{1}} \ell_{2}^{+} \nu_{\ell_{2}} \, jj$ & \\
\bf 260 & $p \overset{\mbox{\tiny{(--)}}}{p} \to W^{-}W^{-} \,  jj\to \ell_{1}^{-} \bar{\nu}_{\ell_{1}} \ell_{2}^{-} \bar{\nu}_{\ell_{2}} \, jj$ & \\
&\\
\hline
&\\
\bf 300 & $p \overset{\mbox{\tiny{(--)}}}{p} \to W^{+}W^{-} \to \ell_{1}^{+} \nu_{\ell_{1}} \ell_{2}^{-}\bar{\nu}_{\ell_{2}} $ & anomalous $VVV$ and $HVV$ couplings\\
\bf 310 & $p \overset{\mbox{\tiny{(--)}}}{p} \to W^{+}Z \to  \ell_{1}^{+} \nu_{\ell_1}  \ell_{2}^{+} \ell_{2}^{-} $ & \ldelim \} {2}{0.8cm} \multirow{2}{*}{anomalous $VVV$ couplings}\\
\bf 320 & $p \overset{\mbox{\tiny{(--)}}}{p} \to W^{-}Z \to \ell_{1}^{-} \bar{\nu}_{\ell_{1}}  \ell_{2}^{+} \ell_{2}^{-} $ & \\
\bf 330 & $p \overset{\mbox{\tiny{(--)}}}{p} \to ZZ \to \ell_{1}^{-} \ell_{1}^{+}  \ell_{2}^{-} \ell_{2}^{+} $ & anomalous $HVV$ couplings\\
\bf 340 & $p \overset{\mbox{\tiny{(--)}}}{p} \to W^{+}\gamma \to \ell_{1}^{+} \nu_{\ell_1} \gamma $ & \ldelim \} {2}{0.8cm} \multirow{2}{*}{anomalous $VVV$ couplings}\\
\bf 350 & $p \overset{\mbox{\tiny{(--)}}}{p} \to W^{-}\gamma \to \ell_{1}^{-} \bar{\nu}_{\ell_1} \gamma $ & \\
\bf 360 & $p \overset{\mbox{\tiny{(--)}}}{p} \to Z\gamma \to \ell_{1}^{-} \ell_{1}^{+}  \gamma $ & \ldelim \} {2}{0.8cm} \multirow{2}{*}{anomalous $HVV$ couplings}\\
\bf 370 & $p \overset{\mbox{\tiny{(--)}}}{p} \to \gamma\gamma $ & \\
&\\
\hline
\end{tabular}
\label{tab:prc3}
\end{center}
\end{table}

\newpage
\begin{table}[htb!]
\newcommand{\lstrut}{{$\strut\atop\strut$}}
\begin{center}
\small
\begin{tabular}{c|l|l}
\hline
&\\
\textsc{ProcId} & \textsc{Process} & \textsc{BSM}  \\
&\\
\hline
&\\
\bf 400 & $p \overset{\mbox{\tiny{(--)}}}{p} \to W^{+}W^{-}Z \to \ell_{1}^{+}\nu_{\ell_{1}} \ell_{2}^{-} \bar{\nu}_{\ell_{2}} 
\ell_{3}^{+} \ell_{3}^{-} $ & \ldelim \} {6}{0.8cm} \multirow{6}{*}{\begin{parbox}{3.65cm}{anomalous couplings, Kaluza-Klein models}\end{parbox}}\\
\bf 410 & $p \overset{\mbox{\tiny{(--)}}}{p} \to ZZW^{+} \to  \ell_{1}^{+} \ell_{1}^{-}  \ell_{2}^{+} \ell_{2}^{-} 
 \ell_{3}^{+} \nu_{\ell_{3}} $ & \\
\bf 420 & $p \overset{\mbox{\tiny{(--)}}}{p} \to ZZW^{-} \to \ell_{1}^{+} \ell_{1}^{-}  \ell_{2}^{+} \ell_{2}^{-} 
 \ell_{3}^{-}  \bar{\nu}_{\ell_{3}}$ & \\
\bf 430 & $p \overset{\mbox{\tiny{(--)}}}{p} \to W^{+}W^{-}W^{+} \to \ell_{1}^{+}\nu_{\ell_1} \ell_{2}^{-}
\bar{\nu}_{\ell_2} \ell_{3}^{+}\nu_{\ell_{3}}$ & \\
\bf 440 & $p \overset{\mbox{\tiny{(--)}}}{p} \to W^{-}W^{+}W^{-} \to \ell_{1}^{-} \bar{\nu}_{\ell_1}\ell_{2}^{+}\nu_{\ell_2}
\ell_{3}^{-} \bar{\nu}_{\ell_{3}} $ & \\
\bf 450 & $p \overset{\mbox{\tiny{(--)}}}{p} \to ZZZ \to \ell_{1}^{-} \ell_{1}^{+} \ell_{2}^{-}
\ell_{2}^{+} \ell_{3}^{-} \ell_{3}^{+} $ & \\
\bf 460 & $p \overset{\mbox{\tiny{(--)}}}{p} \to W^{-}W^{+} \gamma \to \ell_{1}^{-} \bar{\nu}_{\ell_1}
\ell_{2}^{+}\nu_{\ell_2} \gamma$ & \\
\bf 470 & $p \overset{\mbox{\tiny{(--)}}}{p} \to Z Z \gamma \to \ell_{1}^{-} \ell_{1}^{+} \ell_{2}^{-}
\ell_{2}^{+} \gamma$ & \\
\bf 480 & $p \overset{\mbox{\tiny{(--)}}}{p} \to W^{+} Z \gamma \to \ell_{1}^{+}\nu_{\ell_1} \ell_{2}^{-}
\ell_{2}^{+} \gamma$ & \\
\bf 490 & $p \overset{\mbox{\tiny{(--)}}}{p} \to W^{-} Z \gamma \to \ell_{1}^{-} \bar{\nu}_{\ell_1} \ell_{2}^{-}
\ell_{2}^{+} \gamma$ & \\
\bf 500 & $p \overset{\mbox{\tiny{(--)}}}{p} \to W^{+} \gamma \gamma \to {\ell}^{+}\nu_{\ell} 
\gamma \gamma$ & \\
\bf 510 & $p \overset{\mbox{\tiny{(--)}}}{p} \to W^{-} \gamma \gamma \to {\ell}^{-} \bar{\nu}_{\ell} 
\gamma \gamma$ & \\
\bf 520 & $p \overset{\mbox{\tiny{(--)}}}{p} \to Z \gamma \gamma \to {\ell}^{-} {\ell}^{+} 
\gamma \gamma$ & \\
\bf 521 & $p \overset{\mbox{\tiny{(--)}}}{p} \to Z \gamma \gamma \to \nu_{\ell} \bar{\nu}_{\ell} 
\gamma \gamma$ & \\
\bf 530 & $p \overset{\mbox{\tiny{(--)}}}{p} \to \gamma \gamma \gamma $ & \\
&\\
\hline
&\\
\bf 610 & $p \overset{\mbox{\tiny{(--)}}}{p}  \to W^{-} \gamma j \to \ell^{-} \bar \nu_{\ell} \gamma j $ & \ldelim \} {5}{0.8cm} \multirow{5}{*}{anomalous couplings}\\
\bf 620 & $p \overset{\mbox{\tiny{(--)}}}{p}  \to W^{+} \gamma j  \to \ell^{+} \nu_{\ell} \gamma j $ &\\
\bf 630 & $p \overset{\mbox{\tiny{(--)}}}{p}  \to W^{-} Z j \to \ell_{1}^{-} \bar \nu_{\ell_1} \ell_{2}^{-}
\ell_{2}^{+} j$ & \\
\bf 640 & $p \overset{\mbox{\tiny{(--)}}}{p}  \to W^{+} Z j \to  \ell_{1}^{+}\nu_{\ell_1} \ell_{2}^{-}
\ell_{2}^{+}j $ & \\
&\\
\hline
&\\
\bf 800 & $p \overset{\mbox{\tiny{(--)}}}{p}  \to W^{+} \gamma \gamma j  \to \ell^{+} \nu_{\ell} \gamma \gamma j $ &\\
\bf 810 & $p \overset{\mbox{\tiny{(--)}}}{p}  \to W^{-} \gamma \gamma j \to \ell^{-} \bar \nu_{\ell} \gamma \gamma j $ & \\
&\\
\hline
\end{tabular}
\end{center}
\end{table}

\newpage
The processes accessed via the executable {\tt ggflo} are given below.

\begin{table}[htb!]
\newcommand{\lstrut}{{$\strut\atop\strut$}}
\begin{center}
\small
\begin{tabular}{c|l|l}
\hline
&\\
\textsc{ProcId} & \textsc{Process} & \textsc{Bsm}\\
&\\
\hline
&\\
\bf 4100 & $p \overset{\mbox{\tiny{(--)}}}{p} \to H \, jj $ & MSSM, general 2HDM\\
\bf 4101 & $p \overset{\mbox{\tiny{(--)}}}{p} \to H \, jj\to \gamma\gamma \, jj$ & \ldelim \} {9}{0.8cm} \multirow{9}{*}{MSSM}\\
\bf 4102 & $p \overset{\mbox{\tiny{(--)}}}{p} \to H \, jj\to \mu^+\mu^- \, jj$ &\\
\bf 4103 & $p \overset{\mbox{\tiny{(--)}}}{p} \to H \, jj\to \tau^+\tau^- \, jj$ &\\
\bf 4104 & $p \overset{\mbox{\tiny{(--)}}}{p} \to H \, jj\to b\bar{b} \, jj$ & \\
\bf 4105 & $p \overset{\mbox{\tiny{(--)}}}{p} \to H \, jj\to W^{+}W^{-} \, jj\to \ell_{1}^+\nu_{\ell_{1}} \ell_{2}^- 
\bar{\nu}_{\ell_{2}} \,jj$ & \\
\bf 4106 & $p \overset{\mbox{\tiny{(--)}}}{p} \to H \, jj\to ZZ \, jj\to \ell_{1}^+ \ell_{1}^- \ell_{2}^+ 
\ell_{2}^- \,jj$ & \\
\bf 4107 & $p \overset{\mbox{\tiny{(--)}}}{p} \to H \, jj\to ZZ \, jj\to \ell_{1}^+ \ell_{1}^- \nu_{\ell_{2}}  
\bar{\nu}_{\ell_{2}} \,jj$ & \\
&\\
\hline
&\\
\bf 4300 & $gg \to W^{+}W^{-} \to \ell_{1}^{+} \nu_{\ell_{1}} \ell_{2}^{-}\bar{\nu}_{\ell_{2}} $ & \ldelim \} {4}{0.8cm} \multirow{4}{*}{anomalous $HVV$ couplings}\\
\bf 4330 & $gg \to ZZ \to \ell_{1}^{-} \ell_{1}^{+}  \ell_{2}^{-} \ell_{2}^{+} $ & \\
\bf 4360 & $gg \to Z\gamma \to \ell_{1}^{-} \ell_{1}^{+}  \gamma $ & \\
\bf 4370 & $gg \to \gamma\gamma $ & \\
&\\
\hline
\end{tabular}
\end{center}
\end{table}


\begingroup\begin{thebibliography}{10}

\bibitem{Arnold:2008rz}
  K.~Arnold, M.~Bahr, G.~Bozzi {\it et al.},
  ``VBFNLO: A parton level Monte Carlo for processes with electroweak bosons'',
  {\em Comput.\ Phys.\ Commun.}  {\bf 180 } (2009)  1661-1670, \href{http://arxiv.org/abs/0811.4559}{{\tt arXiv:0811.4559}}.

\bibitem{Arnold:2011wj}
  K.~Arnold, J.~Bellm, G.~Bozzi {\it et al.},
  ``VBFNLO: A parton level Monte Carlo for processes with electroweak bosons -- Manual for Version 2.5.0,''
  \href{http://arxiv.org/abs/1107.4038}{{\tt arXiv:1107.4038}}.

\bibitem{Jager:2009xx}
  B.~Jager, C.~Oleari and D.~Zeppenfeld,
  ``Next-to-leading order QCD corrections to $W^+ W^+ jj$ and $W^- W^- jj$ production via weak-boson fusion,''
  {\em Phys.\ Rev.} {\bf D80} (2009) 034022,
\href{http://arxiv.org/abs/0907.0580}{{\tt arXiv:0907.0580}}.

\bibitem{Campanario:2011ud}
  F.~Campanario, C.~Englert, M.~Rauch and D.~Zeppenfeld,
  ``Precise predictions for $W \gamma \gamma$+jet production at hadron colliders,''
  {\em Phys.\ Lett.\ B} {\bf 704} (2011) 515,
  \href{http://www.arXiv.org/abs/1106.4009}{{\tt arXiv:1106.4009}}.
  
%\cite{Campanario:2011cs}
\bibitem{Campanario:2011cs}
  F.~Campanario,
  ``Towards $pp \rightarrow VVjj$ at NLO QCD: Bosonic contributions to triple vector boson production plus jet,''
  {\em JHEP} {\bf 1110} (2011) 070,
  \href{http://www.arXiv.org/abs/arXiv:1105.0920}{{\tt arXiv:1105.0920}}.
  %%CITATION = ARXIV:1105.0920;%%

\bibitem{Jager:2010aj}
  B.~Jager,
  ``Next-to-leading order QCD corrections to photon production via weak-boson
  fusion'',
  {\em Phys.\ Rev.} {\bf D81} (2010) 114016,
 \href{http://arxiv.org/abs/1004.0825}{{\tt arXiv:1004.0825}}. 

\bibitem{frank}
J.~Frank, ``Spin-2 Resonances in Vector Boson Fusion Processes at the LHC'', {Diploma Thesis, ITP Karlsruhe 2011}, {{\tt http://www-itp.particle.uni-karlsruhe.de/diplomatheses.en.shtml}}.

\bibitem{Passarino:1978jh}
  G.~Passarino and M.~J.~G.~Veltman,
  `{`One Loop Corrections for $e^+e^-$ Annihilation into $\mu^+\mu^-$ in the Weinberg
  Model}'',
  {\em Nucl.\ Phys.}  {\bf B160} (1979) 151.

\bibitem{Hagiwara:1993qt}
K.~Hagiwara, R.~Szalapski and D.~Zeppenfeld, ``{Anomalous Higgs boson
  production and decay}'', {\em Phys.\ Lett.} {\bf B318} (1993) 155,
\href{http://www.arXiv.org/abs/hep-ph/9308347}{{\tt hep-ph/9308347}}.

\bibitem{Hagiwara:1993ck}
K.~Hagiwara, S.~Ishihara, R.~Szalapski and D.~Zeppenfeld, ``{Low-energy effects
  of new interactions in the electroweak boson sector}'', {\em Phys.\ Rev.} {\bf
  D48} (1993) 2182.

\bibitem{Achard:2004kn}
{L3} Collaboration, ``{Search for anomalous couplings in
  the Higgs sector at LEP}'', {\em Phys.\ Lett.} {\bf B589} (2004) 89,
\href{http://www.arXiv.org/abs/hep-ex/0403037}{{\tt hep-ex/0403037}}.




\end{thebibliography}\endgroup
\end{document}